\documentstyle[12pt,psfig,epsf]{article}
\newcommand{\be}{\begin{equation}}
\newcommand{\ee}{\end{equation}}

\newcommand{\beqa}{\begin{eqnarray}}
\newcommand{\eeqa}{\end{eqnarray}}

\newcommand{\eqref}[1]{(\ref{#1})}

\def\boxit#1{\vbox{\hrule\hbox{\vrule\kern8pt
\vbox{\hbox{\kern8pt}\hbox{\vbox{#1}}\hbox{\kern8pt}}
\kern8pt\vrule}\hrule}}
\def\mathboxit#1{\vbox{\hrule\hbox{\vrule\kern8pt\vbox{\kern8pt
\hbox{$\displaystyle #1$}\kern8pt}\kern8pt\vrule}\hrule}}

\def\IB{\relax\hbox{$\inbar\kern-.3em{\rm B}$}}
\def\IC{\relax\hbox{$\inbar\kern-.3em{\rm C}$}}
\def\ID{\relax\hbox{$\inbar\kern-.3em{\rm D}$}}
\def\IE{\relax\hbox{$\inbar\kern-.3em{\rm E}$}}
\def\IF{\relax\hbox{$\inbar\kern-.3em{\rm F}$}}
\def\IG{\relax\hbox{$\inbar\kern-.3em{\rm G}$}}
\def\IGa{\relax\hbox{${\rm I}\kern-.18em\Gamma$}}
\def\IH{\relax{\rm I\kern-.18em H}}
\def\IK{\relax{\rm I\kern-.18em K}}
\def\IL{\relax{\rm I\kern-.18em L}}
\def\IP{\relax{\rm I\kern-.18em P}}
\def\IR{\relax{\rm I\kern-.18em R}}
\def\IZ{\relax\ifmmode\mathchoice
{\hbox{\cmss Z\kern-.4em Z}}{\hbox{\cmss Z\kern-.4em Z}}
{\lower.9pt\hbox{\cmsss Z\kern-.4em Z}} {\lower1.2pt\hbox{\cmsss
Z\kern-.4em Z}}\else{\cmss Z\kern-.4em Z}\fi}

\def\II{\relax{\rm I\kern-.18em I}}

\def\CE {{\cal E}}

\def\CR {{\cal R}}
\def\CS {{\cal S}}

\begin{document}

\hfill CERN-PH-TH/2004-163

\hfill NRCPS-HE-2004-18

\vspace{5cm}
\begin{center}
{\LARGE Tensionless Strings\\
Vertex Operator for Fixed Helicity States

}

\vspace{2cm}
{\sl  G. Savvidy$ $\footnote{\footnotesize\it E-mail: georgios.savvidis@cern.ch} \\
\bigskip
{\sl   CERN Theory Division, CH-1211 Geneva 23, Switzerland}\\
{\sl   Demokritos National Research Center,GR-15310 Athens, Greece}

}
\end{center}
\vspace{60pt}

\centerline{{\bf Abstract}}

\vspace{12pt}

\noindent
The tensionless string theory with perimeter action has pure massless spectrum of
higher-spin gauge fields. The multiplicity of these massless states grows linearly.
It is therefore much less
compared with the standard string theory
and is larger compared with the field theory models of the Yang-Mills type.
It is important to define nontrivial interaction between
infinite amount of massless particles of the perimeter string theory.
The appropriate vertex operators were defined recently and
I study the lowest order vertex operators
and the corresponding scattering amplitudes in tree approximation.
I emphasize the special importance of the vertex operator for fixed helicity states.


\newpage

\pagestyle{plain}

\section{{\it Introduction}}

A string model which is based on the concept of the surface
perimeter\footnote{For the perimeter action
$S_P = (m/4\pi) \int d \tau d \sigma \sqrt{\left(\partial^{2}
X \right)^{2}}$ we shall use physical units in which m=1.}
was suggested in  \cite{Savvidy:dv,Savvidy:2003fx,geo}.
At the classical level the model is {\it tensionless} \cite{geo}
and has a pure massless spectrum of infinitely many integer spin fields
\cite{Savvidy:dv,Savvidy:2003fx}.
It was demonstrated in \cite{Savvidy:2003fx,Antoniadis:2004uh}
that unphysical states are eliminated by the Virasoro
and additional Abelian constraints appearing in the model. The analysis of the
first three lower levels of the physical Fock space ${{\cal F}}$ shows that the
fixed helicity states ${{\cal F}}_{0}$ and the first two excited
states ${{\cal F}}_{1,2}$ are well defined and have no negative norm
waves.  The level ${{\cal F}}_{0}$ of fixed helicity states is infinitely degenerate
and contains massless particles of increasing tensor structure
$A^{\mu_1 ,..., \mu_s}(k)$, where $s=1,2,...$, while the first and the
second levels ${{\cal F}}_{1,2}$ are physical null states
\footnote{In Appendix I discuss the origin of this phenomena at the classical level.}.
It was conjectured in  \cite{Savvidy:2003fx} that all excited
states ${{\cal F}}_{n}$ ~($n \neq 0$) represent physical null states
realizing continuous spin representations of the Poincar\'e group and
define large gauge transformation of the fixed helicity states ${{\cal F}}_{0}$.

The next important issue is connected with the interaction
of these massless fields. The above analysis of the physical Fock space imposes
strong restriction on the possible form of interactions. Indeed only the subspace
${{\cal F}}_{0}$ represents the propagating physical fields, therefore   {\it
while introducing interaction into the perimeter string model it is
important to demonstrate that nontrivial transitions amplitudes
are only between fixed helicity states ${{\cal F}}_{0}$}.

The solution of this problem may shed some new light to the old problem
of the existence of consistent interaction of massless particles of higher
spin fields in flat space-time \cite{pauli,singh,fronsdal,Bengtsson:1983pg}.
The recent attempt to construct the appropriate
vertex operators for the perimeter string model was made in \cite{alvarez}.

The general form of the suggested vertex operators is given by the formula \cite{alvarez}:
\be\label{highordervertex}
U^{\mu_1 \tilde{\mu}_1...,  ...\mu_j \tilde{\mu}_j}_{k,\pi}(\zeta)=
:\partial^{m_1}_{\zeta} X^{\mu_1}~
\partial^{\tilde{m}_1}_{\bar{\zeta}} X^{\tilde{\mu}_1}~...~
~ ...\partial^{m_j}_{\zeta} \Pi^{\mu_j}~\partial^{\tilde{m}_j}_{\bar{\zeta}} \Pi^{\tilde{\mu}_j}~
e^{~i k \cdot X(\zeta) +i\pi \cdot \Pi(\zeta)~}:
\ee
where the canonically conjugate operators are \cite{Savvidy:dv} :
$
X^{\mu}= {1\over 2}(X^{\mu}_{L}(\zeta^+) + X^{\mu}_{R}(\zeta^-))
$
\beqa\label{decom}
X^{\mu}_{L} = x^{\mu} +
\pi^{\mu}\zeta^{+} +
i\sum_{n \neq 0}  {1\over n }~ \beta^{\mu}_{n} e^{-in\zeta^{+}},\nonumber\\
X^{\mu}_{R} =   x^{\mu}
+  \pi^{\mu}\zeta^{-} +
i\sum_{n \neq 0}  {1\over n }~ \tilde{\beta}^{\mu}_{n} e^{-in\zeta^{-}},
\eeqa
and $P^{\mu}_{\tau}=\partial_{\tau}\Pi^{\mu}$,~~where~
$\Pi^{\mu}  = {1\over 2}(\Pi^{\mu}_{L}(\zeta^{+})
+ \Pi^{\mu}_{R}(\zeta^{-}))
$
\beqa\label{decom1}
\Pi^{\mu}_{L}=   e^{\mu} +  k^{\mu}\zeta^{+} +
i\sum_{n \neq 0}  {1\over n }~ \alpha^{\mu}_{n} e^{-in\zeta^{+}},\nonumber\\
\Pi^{\mu}_{R}=   e^{\mu} + k^{\mu}\zeta^{-} +
i\sum_{n \neq 0}  {1\over n }~ \tilde{\alpha}^{\mu}_{n} e^{-in\zeta^{-}}.
\eeqa
The basic commutator is of the form
$
[X^{\mu}_{L,R}(\zeta^{\pm}),\partial_{\pm}
\Pi^{\nu}_{L,R}(\zeta^{'\pm})]= 2 \pi i\eta^{\mu\nu} \delta
(\zeta^{\pm} - \zeta^{'\pm})
$
and the following commutator relations hold for the oscillators
$\alpha_n$,$\beta_n$
$$
[e^{\mu}, \pi^{\nu}]=[x^{\mu}, k^{\nu}] =  i\eta^{\mu\nu},~~~[\alpha^{\mu}_{n},
\beta^{\nu}_{l}]= n~\eta^{\mu\nu}\delta_{n+l,0},
$$
where~ $\alpha^{\mu}_{0} \equiv k^{\mu},~~\beta^{\mu}_{0}\equiv
\pi^{\mu}$~~is a pair of momentum operators.
The momentum $k_{\mu}$ is the
standard space-time momentum operator, the $\pi_{\mu}$ is a new
momentum operator conjugate to the polarization vector coordinate $e_{\mu}~$~ \cite{Savvidy:dv}.

Using the world-sheet energy momentum operator \cite{Savvidy:dv}
$T(\zeta)=-:\partial_{\zeta} X \cdot \partial_{\zeta}
\Pi:$ one can compute the anomalous dimension of the vertex
operators in (\ref{highordervertex}) \cite{alvarez}:
\be\label{conformaldimensions}
\Delta = 2(k \cdot \pi) + m_1 + \tilde{m}_1 +...
...+ m_j+ \tilde{m}_j.
\ee
It must be equal to $2$ in order to describe
emission of physical states, therefore the corresponding poles are at the points
\be\label{conformalinvarianceconditions}
(k \cdot \pi) = 1 - (m_1 + \tilde{m}_1 +...
...+ m_j+ \tilde{m}_j)/2~~,
\ee
(the conformal spin should be equal to
zero, therefore $m_1 +...+m_j = \tilde{m}_1 +...+\tilde{m}_j$).

The two lowest order vertex operators of conformal dimension two
are the $V_{k,\pi}$ operator:
$$
V_{k,\pi}(\zeta) = : e^{~i k \cdot X(\zeta) +i\pi \cdot \Pi(\zeta)~}:~~~~~~~~
~~~~~~~~(k \cdot \pi)=1
$$
and the $U_{k,\pi}$ operators of the type $m = \tilde{m} =1$:
$$
U^{\mu\nu}_{k, \pi}(\zeta) =
: \partial_{\zeta} X^{\mu} ~\partial_{\bar{\zeta}}\Pi^{\nu}~
~e^{~i k \cdot X(\zeta) +i\pi \cdot \Pi(\zeta)~}:~~~~
~~~(k \cdot \pi)=0.
$$
The $U_{k,\pi}$ operators are of the essential importance, because for them
$(k \cdot \pi)=0$  and they
create fixed helicity states ${{\cal F}}_{0}$ \cite{Savvidy:2003fx}. Indeed the square
of the Pauli-Lubanski vector of the Poincar\'e group is $W= (k \cdot \pi)^2 =  \Xi^2$
and defines fixed helicity states, when $W=0$ and
continuous spin representations-CSR, when $W \neq 0$
\cite{wigner,brink,Savvidy:dv,Savvidy:2003fx}.

The aim of this article is to compute the residues of
different poles in the scattering amplitudes defined by the above
vertex operators and to derive consistency conditions under which
all transition amplitudes to the CSR states from fixed helicity
states are vanishing. In the
next sections I shall present necessary material from \cite{alvarez} and shall derive
the consistency conditions under which the residue of the transition amplitude from fixed helicity
state $U_{k,\pi}(0)|0>$ ~ to the CSR state $V_{k,\pi}(0)|0>$~  is equal to zero.

\section{{\it The Lowest Order  Vertex Operators}}

Let us consider the lowest order vertex operators in (\ref{highordervertex}), which have been
defined as $V_{k,\pi}$ and $U_{k,\pi}$ operators in \cite{alvarez}.
The most lower order vertex operator $V_{k,\pi}$ has the form:
\be
V_{k,\pi}(\zeta) = : e^{~i k \cdot X(\zeta) +i\pi \cdot \Pi(\zeta)~}:~~~~~~~~(k \cdot \pi)=1
\ee
and at the next level there are three relevant operators with
$m = \tilde{m} =1$ \cite{alvarez}:
\beqa\label{lightconegroundstatevertices}
U^{\mu\nu}_{XX}(k, \pi,\zeta) &=&
:(\partial_{\zeta} X^{\mu}~ \partial_{\bar{\zeta}} X^{\nu} +
\partial_{\bar{\zeta}} X^{\mu}~ \partial_{\zeta} X^{\nu} )~
e^{~i k \cdot X(\zeta) +i\pi \cdot \Pi(\zeta)~}:\nonumber\\
U^{\mu\nu}_{X\Pi}(k, \pi,\zeta) &=&
: (\partial_{\zeta} X^{\mu} ~\partial_{\bar{\zeta}}\Pi^{\nu} +
\partial_{\bar{\zeta}} X^{\mu} ~\partial_{\zeta}
\Pi^{\nu})~e^{~i k \cdot X(\zeta) +i\pi \cdot \Pi(\zeta)~}:~~~~~~~~(k \cdot \pi)=0\nonumber\\
U^{\mu\nu}_{\Pi\Pi}(k, \pi,\zeta) &=&
: (\partial_{\zeta} \Pi^{\mu} ~\partial_{\bar{\zeta}}\Pi^{\nu} +
\partial_{\bar{\zeta}}\Pi^{\mu} ~\partial_{\zeta} \Pi^{\nu})~
e^{~i k \cdot X(\zeta) +i\pi \cdot \Pi(\zeta)~}: ~~,
\eeqa
so that the general form of the fixed helicity state vertex operator
$U_{k,\pi}$ is given by the linear
combination of these operators
\be\label{generalformofgroundstatevertex}
U_{k,\pi}  = \omega_{\mu\nu } U^{\mu\nu}_{XX}+2\varphi_{\mu\nu } U^{\mu\nu}_{X\Pi}
+\chi_{\mu\nu } U^{\mu\nu}_{\Pi\Pi},
\ee
where $\omega(k,\pi),\varphi(k,\pi)$ and $\chi(k,\pi)$  are polarization
tensors, or in our case it will be more precise to call them coupling constants.

There are two types of constraints which are imposed on the physical
states and transition amplitudes of the perimeter model: these are the
standard Virasoro constraints and the additional Abelian constraints \cite{Savvidy:dv}.
The general vertex operators (\ref{highordervertex}) are already built in a way to
fulfill the conformal invariance of the amplitudes (\ref{conformalinvarianceconditions}).
All correlation functions
should also be restricted to fulfill the Abelian $\Theta = \Pi^2 -1 =0$ constraint.
As a result some of the vertex operators (\ref{highordervertex}) are excluded as unphysical.
This happens with the lowest order vertex operator $V_{k,\pi}$
\cite{alvarez}. For the vertex operator $V_{k,\pi}$
the Abelian constraint takes the form:
\be\label{wordidentity}
<:\Pi^2(\eta): :e^{~i k \cdot X(\zeta) +i\pi \cdot \Pi(\zeta)~}:>=
<:e^{~i k \cdot X(\zeta) +i\pi \cdot \Pi(\zeta)~}:>.
\ee
and reduces to the equation
$$
-k^2 ~\ln^2 \vert \eta - \zeta \vert =  \left\{ \begin{array}{l}
  0 ~~if ~~k_{\mu} \neq 0~~~ or~~~ \pi_{\mu} \neq 0\\
  1 ~~if ~~k_{\mu} =0~~ and~~ \pi_{\mu} = 0
\end{array} \right\}.
$$
To eliminate the world-sheet coordinate dependence on the l.h.s. we have to
impose at least the massless condition
\be\label{masslesscondition}
k^2 =0,
\ee
which includes the point $k_{\mu} = 0$. Therefore the constraint is
fulfilled for nonzero $k_{\mu}$ and breaks down at the point $k_{\mu} = 0$.
Thus the constraint (\ref{wordidentity})
can not be fulfilled in the deep infrared region where $k_{\mu}=0$
and  the vertex operator $V_{k,\pi}$ should be
abandoned as unphysical.

The Abelian constraint $\Theta$ should be
imposed on the higher order vertex
operators $U_{k,\pi}$ as well
$
<:\Pi^2(\eta): U_{k,\pi}(\zeta)>~=~
<U_{k,\pi}(\zeta)>
$ and this condition is fulfilled if $k^2 =0$ and
\be\label{traceofomega}
Tr \omega =0.
\ee
Notice that now the V.E.V. on the r.h.s.
is equals to zero $<U_{k,\pi}(\zeta)> =0$.

Because the states created by the operator $V_{k,\pi}$ are
unphysical the consistency of the model requires that the transition amplitudes
from fixed helicity
state $U_{k,\pi}(0)|0>$ ~ to the CSR state $V_{k,\pi}(0)|0>$~  must be equal to zero.

Let us therefore consider the vertex operators $U_{k, \pi}$ in more details.
These operators have the following operator product expansions \cite{alvarez} with the
stress tensor $T(\eta)=-:\partial_{\eta}X\cdot\partial_{\eta}\Pi$ \cite{Savvidy:dv}:
\beqa\label{productwithstresstensor}
T(\eta) U^{\mu\nu}_{XX}(\zeta) &=& -{1\over (\eta - \zeta)^3}
:(i\pi^{\mu}\partial_{\bar{\zeta}} X^{\nu} + i\pi^{\nu}\partial_{\bar{\zeta}} X^{\mu})
e^{i k \cdot X +i\pi \cdot \Pi}:~\nonumber\\
&~&~~~~~~~~~~~~~~~~~~~~~~~~~~~~~~~~~~~~~~~~~~~~~~
+({ k\pi  + 1\over (\eta - \zeta)^2} + {1\over (\eta - \zeta)}
\partial_{\zeta} ) U^{\mu\nu}_{XX}(\zeta), \nonumber\\
T(\eta) U^{\mu\nu}_{X\Pi}(\zeta)&=&-{1\over (\eta - \zeta)^3}
:(i\pi^{\mu}\partial_{\bar{\zeta}} \Pi^{\nu} +
ik^{\nu}\partial_{\bar{\zeta}} X^{\mu})~e^{~i k \cdot X +i\pi \cdot \Pi~}:~\nonumber\\
&~&~~~~~~~~~~~~~~~~~~~~~~~~~~~~~~~~~~~~~~~~~~~~~~+({ k\pi  + 1\over (\eta - \zeta)^2}~
  + {1\over (\eta - \zeta)}
\partial_{\zeta} )U^{\mu\nu}_{X\Pi}(\zeta),\nonumber\\
T(\eta) U^{\mu\nu}_{\Pi\Pi}(\zeta) &=&-{1\over (\eta - \zeta)^3}
:(ik^{\mu}\partial_{\bar{\zeta}} \Pi^{\nu}+ik^{\nu}\partial_{\bar{\zeta}} \Pi^{\mu})
e^{i k \cdot X +i\pi \cdot \Pi}:~\nonumber\\
&~&~~~~~~~~~~~~~~~~~~~~~~~~~~~~~~~~~~~~~~~~~~~~~~
+({ k\pi  + 1\over (\eta - \zeta)^2}  + {1\over (\eta - \zeta)}
\partial_{\zeta}) U^{\mu\nu}_{\Pi\Pi}(\zeta).\nonumber
\eeqa

The linear combination $U_{k,\pi}=\omega U_{XX}+2\varphi U_{X\Pi}
+\chi U_{\Pi\Pi}$ should be such that the leading singularities
$(\eta - \zeta)^{-3}$ cancel, making the operator product characteristic
to the conformal dimension-one primary fields. This leads to the equations
\be\label{polarizationequation1}
\pi^{\mu} \omega_{\mu\nu } + \varphi_{\nu \mu} k^{\mu} =0,~~~~
 \pi^{\mu}\varphi_{\mu\nu } + k^{\mu} \chi_{\mu\nu } =0,~~~(k \cdot \pi)=0.
\ee
An analogous expansion holds between $\bar{T}(\bar{\zeta})$ and the three vertices,
leading to the equations
\be\label{polarizationequation2}
\omega_{\mu\nu } \pi^{\nu} + \varphi_{\mu\nu } k^{\nu} =0,~~~
\pi^{\mu}\varphi_{\mu\nu }  + \chi_{\nu \mu} k^{\mu}=0,~~~(k \cdot \pi)=0.
\ee
Because the tensors $\omega$ and $\chi$ are symmetric, both equations
simply coincide.

The equations (\ref{traceofomega}),(\ref{polarizationequation1}) and
(\ref{polarizationequation2}) are not the only ones which are imposed
on the tensors $\omega$, $\varphi$  and $\chi$. Indeed, as we just explained
the lowest order vertex operator $V_{k,\pi}$ is unphysical and the quantum mechanical consistency
requires that the scattering of two states created by the operator $U_{k,\pi}$
into the state created by $V_{k,\pi}$ must vanish. In the next section we shall derive
this condition.

\section{{\it The zero residue condition}}

Suppose that we are scattering $n$ states which are created by the operators
$U_{k,\pi}$. To find the poles and the corresponding residues of the
intermediate states, which appear in the tree diagram, we have to compute
the operator product of the form
\beqa\label{operatorproductofUs}
U_{k_1 , \pi_1}(\zeta_1)~ U_{k_2 , \pi_2}(\zeta_2) ~ \simeq~
\CR \CE \CS_0 ~~
\vert \zeta_1 - \zeta_2 \vert^{2(k_1 \pi_2 + k_2 \pi_1) -4}~~
V_{k_1 +k_2 , \pi_1 +\pi_2}(\zeta_2) +~~~~~\nonumber\\
\CR \CE \CS_1 ~~  \vert \zeta_1 - \zeta_2 \vert^{2(k_1 \pi_2 + k_2 \pi_1) -2} ~~
U_{k_1 +k_2 , \pi_1 +\pi_2}(\zeta_2) +.....,
\eeqa
where in general the lowest order vertex operator $V_{k,\pi}$ has appeared on the r.h.s.
together with higher order ones. The corresponding residues are $\CR \CE \CS_0$,~
$\CR \CE \CS_1$ and so on. To find this residues we have to compute the
two-point correlation functions when $k_1 \pi_1=k_2 \pi_2=0$.
In the following we shall not discuss the
contribution of the higher order operators on the r.h.s., but should
mention that their contribution should also vanish. {\it Thus, the only transitions
which are permitted are between $U_{k,\pi}$ states of fixed helicity.}

The two point correlation function of the vertex operator
$U_{k , \pi}  = \omega U_{XX} +2\varphi U_{X\Pi} +\chi U_{\Pi\Pi} $ contains six cross terms
which have the form \cite{alvarez}:
\beqa\label{paircorrelations}
<U^{\mu_1 \nu_1}_{XX} 
U^{\mu_2 \nu_2}_{XX}
>~
 = ~\pi^{\mu_1}_{2}\pi^{\nu_1}_{2}
\pi^{\mu_2}_{1} \pi^{\nu_2}_{1}~
\vert \zeta_1 - \zeta_2 \vert^{2(k_1 \pi_2 + k_2 \pi_1) -4},~~~~~~~~~~~~~~~~~~~~~~~~~~~~~~~~\nonumber
\eeqa
\beqa
<U^{\mu_1 \nu_1}_{X\Pi} 
U^{\mu_2 \nu_2}_{X\Pi}
>~
= 2~\{\eta^{\mu_1 \nu_2}~\eta^{\nu_1\mu_2} -  \eta^{\mu_1 \nu_2}
k^{\nu_1}_2 \pi^{\mu_2}_1~- \eta^{\nu_1 \mu_2}
\pi^{\mu_1}_{2} k^{\nu_2}_{1}   +~~~~~~~~~~~~~~~~~~~\nonumber\\
+ 2\pi^{\mu_1}_{2} k^{\nu_1}_{2}
\pi^{\mu_2}_{1}k^{\nu_2}_{1} ~\}~
\vert \zeta_1 - \zeta_2 \vert^{2(k_1 \pi_2 + k_2 \pi_1) -4},\nonumber
\eeqa
\beqa
<U^{\mu_1 \nu_1}_{\Pi\Pi} 
U^{\mu_2 \nu_2}_{\Pi\Pi}
>~ = ~k^{\mu_1}_{2} k^{\nu_1}_{2}  k^{\mu_2}_{1}  k^{\nu_2}_{1}~
\vert \zeta_1 - \zeta_2 \vert^{2(k_1 \pi_2 + k_2 \pi_1) -4},~~~~~~~~~~~~~~~~~~~~~~~~~~~~~~~~~~\nonumber
\eeqa
\beqa
<U^{\mu_1 \nu_1}_{XX} 
U^{\mu_2 \nu_2}_{X\Pi}
>~ =( 2 \pi^{\mu_1}_{2}\pi^{\nu_1}_{2}\pi^{\mu_2}_1
k^{\nu_2}_{1} -\eta^{\nu_1 \nu_2}\pi^{\mu_1}_{2}\pi^{\mu_2}_1
-\eta^{\mu_1 \nu_2}\pi^{\nu_1}_{2}\pi^{\mu_2}_1
)~\vert \zeta_1 - \zeta_2 \vert^{2(k_1 \pi_2 + k_2 \pi_1) -4},~~~~~~~~~~~~\nonumber
\eeqa
\beqa
<U^{\mu_1 \nu_1}_{XX} 
U^{\mu_2 \nu_2}_{\Pi\Pi}
>~=(\eta^{\mu_1 \mu_2}~\eta^{\nu_1\nu_2} - ~\eta^{\mu_1 \mu_2}
\pi^{\nu_1}_2 k^{\nu_2}_1 ~- ~\eta^{\nu_1 \nu_2}
\pi^{\mu_1}_{2} k^{\mu_2}_{1}   +~~~~~~~~~~~~~~\nonumber\\
+\pi^{\mu_1}_{2} k^{\mu_2}_{1}
\pi^{\nu_1}_{2}k^{\nu_2}_{1} )
\vert \zeta_1 - \zeta_2 \vert^{2(k_1 \pi_2 + k_2 \pi_1) -4},\nonumber
\eeqa
\beqa
<U^{\mu_1 \nu_1}_{X\Pi} 
U^{\mu_2 \nu_2}_{\Pi\Pi}
>~=(2\pi^{\mu_1}_{2} k^{\nu_1}_{2}k^{\mu_2}_{1} k^{\nu_2}_{1}
-\eta^{\mu_1 \mu_2}~k^{\nu_1}_{2}k^{\nu_2}_{1}
-\eta^{\mu_1 \nu_2}~k^{\nu_1}_{2}k^{\mu_2}_{1})
\vert \zeta_1 - \zeta_2 \vert^{2(k_1 \pi_2 + k_2 \pi_1) -4}\nonumber
\eeqa
and the full two point correlation function is:
\beqa
<U_{k_1 , \pi_1}, U_{k_2 , \pi_2} > =~~~~~~~~~~~~~~~~~~~~~~~~~~~~~~~~~~~~~~~
\nonumber\\ =<\omega_1 U_{XX}  +2\varphi_1 U_{X\Pi}  +
\chi_1 U_{\Pi\Pi}, ~\omega_2 U_{XX}  +2\varphi_2 U_{X\Pi}  +
\chi_2 U_{\Pi\Pi}>\nonumber\\
= \CR \CE \CS_0 ~~~\vert \zeta_1 - \zeta_2 \vert^{2(k_1 \pi_2 + k_2 \pi_1) -4}.
\eeqa
Using the above pairs of correlation functions (\ref{paircorrelations}) we can get
\beqa\label{residuepolynomial}
\CR \CE \CS_0  =(\pi_2 \omega_1 \pi_2)(\pi_1 \omega_2 \pi_1)+\nonumber\\
2 tr(\varphi_1 \varphi_2) - 2(\varphi_1 k_2) ~(\pi_1 \varphi_2) -
2(\pi_2 \varphi_1) ~(\varphi_2 k_1) +
4 (\pi_2 \varphi_1 k_2)(\pi_1 \varphi_2 k_1)+\nonumber\\
(k_2 \chi_1 k_2)(k_1 \chi_2 k_1)+\nonumber\\
2(\pi_2 \omega_1 \pi_2) (\pi_1 \varphi_2 k_1) - (\pi_2 \omega_1)(\pi_1 \varphi_2) ~ -
(\omega_1 \pi_2 ) ~(\pi_1 \varphi_2)+\nonumber\\
tr(\omega_1 \chi_2) - (\omega_1 \pi_2) ~(\chi_2 k_1 ) -
(\pi_2 \omega_1) ~(k_1 \chi_2 ) +
(\pi_2 \omega_1 \pi_2)(k_1 \chi_2 k_1)+\nonumber\\
2 (\pi_1 \omega_2 \pi_1)(\pi_2 \varphi_1 k_2))
- (\pi_1 \omega_2) ~(\pi_2 \varphi_1 ) -
(\omega_2 \pi_1)~(\pi_2 \varphi_1) +\nonumber\\
2 (\pi_2 \varphi_1 k_2)(k_1 \chi_2 k_1)- (\varphi_1 k_2) ~(\chi_2 k_1 ) -
(\varphi_1 k_2 ) ~(k_1 \chi_2 ) +\nonumber\\
tr(\chi_1 \omega_2) - (\chi_1 k_2) ~(\omega_2 \pi_1 ) -
(k_2 \chi_1) ~(\pi_1 \omega_2 ) +
(k_2 \chi_1 k_2)(\pi_1 \omega_2 \pi_1)+\nonumber\\
2 (k_2 \chi_1 k_2)(\pi_1 \varphi_2 k_1)- (\chi_1 k_2) ~(\varphi_2 k_1 ) -
(k_2 \chi_1  ) ~(\varphi_2 k_1 )
\eeqa
The vertex operator $V_{k_1 +k_2 , \pi_1 +\pi_2}$ should have anomalous
dimension $\Delta$ equal to two (\ref{conformalinvarianceconditions})
$2(k_1 +k_2)\cdot (\pi_1 +\pi_2) =2(k_1 \pi_2 + k_2 \pi_1) = 2$, therefore the
transition amplitude to the state $V_{k,\pi}(0)|0>$ is equal to zero if
\be\label{zeroresiduecondition}
\CR \CE \CS_0 =0,~~~ at~~~ k_1 \pi_2 + k_2 \pi_1 = 1.
\ee
In summary we have the system of equations which define the tensors
$\omega$, $\varphi$  and $\chi$: they are the equations
(\ref{traceofomega}),(\ref{polarizationequation1}) and
(\ref{polarizationequation2}) together with  (\ref{zeroresiduecondition}) and
(\ref{residuepolynomial}).

When the above zero residue condition (\ref{zeroresiduecondition}) is fulfilled
then the three point space-time vertex function will be given by the residue $\CR \CE \CS_1$:
\beqa
U_{k_1 , \pi_1}(\zeta_1)~ U_{k_2 , \pi_2}(\zeta_2)~
\simeq
\CR \CE \CS_1 ~~  \vert \zeta_1 - \zeta_2 \vert^{2(k_1 \pi_2 + k_2 \pi_1) -2} ~~
U_{k_1 +k_2 , \pi_1 +\pi_2}(\zeta_2) +.....
\eeqa
at $\Delta= 2(k_1 +k_2)\cdot ( \pi_1 +\pi_2) +2  =2(k_1 \pi_2 + k_2 \pi_1) +2 = 2$, that is
\be\label{fixedhelicitypole}
k_1 \pi_2 + k_2 \pi_1 =0.
\ee
The general solution of the equations
(\ref{traceofomega}),(\ref{polarizationequation1}),
(\ref{polarizationequation2}) and (\ref{zeroresiduecondition})
has not been found yet, but some particular
two-parameter family of solutions of fixed helicity operators $U^{T}_{k,\pi}$ can be found.
The solution will be presented in the next section.
Unfortunately the scattering of this family of states creates the state which does not
belong to the same family and one
should find more general solution. Nevertheless this exercise can provide
necessary information about the general solution.

\section{{\it Examples of Fixed Helicity Vertex Operators}}

Let us consider the derivatives of the operator $V_{k,\pi}$
$$
U^{L}_{k,\pi} = \partial_{\zeta}\partial_{\bar{\zeta}}V_{k,\pi} = :(i k \cdot \partial_{\zeta}X +
i\pi \cdot \partial_{\zeta}\Pi~)(i k \cdot \partial_{\bar{\zeta}}X +
i\pi \cdot \partial_{\bar{\zeta}}\Pi~)~e^{~i k \cdot X +i\pi \cdot \Pi~}:
$$
If one takes the invariant $(k \cdot \pi)=0$, then the $U^{L}_{k,\pi}$ is a primary field of anomalous dimension two
and is of the same type as $U_{k,\pi}$ in (\ref{lightconegroundstatevertices}).
Comparing it with the general form of the vertex operator $U_{k,\pi}$ in
(\ref{generalformofgroundstatevertex}),
one can conclude that it is the solution of (\ref{polarizationequation1}) and
(\ref{polarizationequation2}) of the form
\be\label{longitudinalvertex}
\omega_{\mu\nu } = a ~k_{\mu}k_{\nu },~~
\varphi_{\mu\nu } = a ~k_{\mu}\pi_{\nu },~~ \chi_{\mu\nu } = a ~\pi_{\mu}\pi_{\nu }.
\ee
This state corresponds to the longitudinal mode and should decouple from
the physical amplitudes. Indeed, suppose that we are scattering two longitudinal
modes which are given by the above tensors.
Then using our two-point correlation function and the expression for the
corresponding residue (\ref{residuepolynomial}) one can find
\beqa
U^{L}_{k_1 , \pi_1}(\zeta_1)~ U^{L}_{k_2 , \pi_2}(\zeta_2) ~ \simeq~
(k_1 \pi_2 + k_2 \pi_1)^2 (k_1 \pi_2 + k_2 \pi_1 -1)^2 ~~~~~~~~~~~~~~~~~~~~~~~~~~~~~~~~\nonumber\\
\vert \zeta_1 - \zeta_2 \vert^{2(k_1 \pi_2 + k_2 \pi_1) -4}
V_{k_1 +k_2 , \pi_1 +\pi_2}(\zeta_2) +.....
\eeqa
Because $V_{k,\pi}$ should have anomalous dimension equal to two
the pole on the r.h.s. is at the point $k_1 \pi_2 + k_2 \pi_1 =1 ~$(\ref{zeroresiduecondition}).
This implies that the residue of the
corresponding pole is equal to zero
$$
\CR \CE \CS_0 = (k_1 \pi_2 + k_2 \pi_1)^2 (k_1 \pi_2 + k_2 \pi_1 -1)^2 =0.
$$

Our aim now is to find new solution of the equations (\ref{polarizationequation1}),
(\ref{polarizationequation2}) with the same property, that is in their operator product
the corresponding residue to the $V_{k,\pi}|0>$ state is equal to zero
(\ref{zeroresiduecondition}).
The condition for that is that the polynomial (\ref{residuepolynomial})
should vanish at the pole (\ref{zeroresiduecondition}): $k_1 \pi_2 + k_2 \pi_1 =
\alpha + \beta =1$.
The scalars $k_1 \pi_2$ and $k_2 \pi_1$ will appear frequently in our
calculations and we shall use special notation for them:
$\alpha = k_1 \pi_2 $ and $\beta = k_2 \pi_1$.

Let us consider the three-parameter solution of equations (\ref{polarizationequation1}),
(\ref{polarizationequation2}) for $U_{k,\pi}$ of the form:
\be
\omega_{\mu\nu } = (1+\varepsilon) ~k_{\mu}k_{\nu },~~
\varphi_{\mu\nu } = (1+\delta) ~k_{\mu}\pi_{\nu },~~ \chi_{\mu\nu } =
(1+\varrho) ~\pi_{\mu}\pi_{\nu },
\ee
where $\varepsilon,\delta$ and $\varrho$ are unknown. Substituting this parameterization
into (\ref{residuepolynomial}) we shall get
\beqa
\CR \CE \CS_0 &=& ((1+\varepsilon)^2 +(1+\varrho)^2) \alpha^2 \beta^2 \nonumber\\
&+&(1+\delta)^2 2 \alpha\beta (1- \alpha - \beta +2 \alpha\beta)\nonumber\\
&+&(1+\varepsilon)(1+\delta)2 \alpha\beta (\alpha^2 + \beta^2 - \alpha - \beta) \nonumber\\
&+&(1+\varepsilon)(1+\varrho)(\alpha^2 + \beta^2 - 2\alpha^3 - 2\beta^3 + \alpha^4 + \beta^4) \nonumber\\
&+&(1+\delta)(1+\varrho)2 \alpha\beta (\alpha^2 + \beta^2 - \alpha - \beta).
\eeqa
At the pole $\alpha + \beta =1$ the corresponding residue should vanish,
thus substituting $\alpha + \beta =1$ into the last equation we shall get
\beqa
\CR \CE \CS_0 &=& [(1+\varepsilon)^2 +(1+\varrho)^2
+ 4(1+\delta)^2 \nonumber\\
&-&4(1+\varepsilon)(1+\delta) \nonumber\\
&+&2(1+\varepsilon)(1+\varrho)\nonumber\\
&-&4(1+\delta)(1+\varrho) ]~~\alpha^2 (1- \alpha^2)\nonumber\\
&=& [(1 + \varepsilon) + (1 + \varrho) - 2(1 +\delta) ]^2 ~~\alpha^2 (1- \alpha^2).
\eeqa
The residue  $\CR \CE \CS_0 $ is equal to zero if
\be
\varepsilon  + \varrho - 2 \delta =0.
\ee
The last equation gives us two-parameter family of operators  with required
property
\beqa
U^{T}_{k , \pi}(\zeta) =(1+\varepsilon):\left(k\cdot \partial_{\zeta} X ~ k\cdot \partial_{\bar{\zeta}} X -
\pi \cdot \partial_{\zeta} \Pi ~ \pi \cdot \partial_{\bar{\zeta}}\Pi \right)~
e^{~i k \cdot X(\zeta) +i\pi \cdot \Pi(\zeta)~}: ~~\nonumber\\
(1+\delta):\left(k\cdot \partial_{\zeta} X ~ \pi\cdot \partial_{\bar{\zeta}} \Pi +
k \cdot \partial_{\bar{\zeta}}X ~ \pi \cdot \partial_{\zeta} \Pi +
2 \pi \cdot \partial_{\zeta} \Pi ~ \pi\cdot \partial_{\bar{\zeta}} \Pi~\right)~
e^{~i k \cdot X(\zeta) +i\pi \cdot \Pi(\zeta)~}:
\eeqa
Thus we can select two independent vertex operators
\be
U^{T_1}_{k , \pi}(\zeta) =:\left( k\cdot \partial_{\zeta} X  ~  k\cdot \partial_{\bar{\zeta}} X  -
 \pi \cdot \partial_{\zeta} \Pi  ~ \pi \cdot \partial_{\bar{\zeta}}\Pi  \right)~
W_{k, \pi}(\zeta):
\ee
and
\beqa
U^{T_2}_{k , \pi}(\zeta) =
:\left( k\cdot \partial  X  ~  \pi\cdot \bar{\partial}  \Pi  +
k \cdot \bar{\partial} X ~ \pi \cdot \partial  \Pi  +
2  \pi \cdot \partial  \Pi  ~ \pi\cdot \bar{\partial}  \Pi ~\right)~
W_{k, \pi}(\zeta):~~,
\eeqa
where for both of them $~(k\cdot \pi) =0$ and
$W_{k, \pi}(\zeta) = \exp{[~i k \cdot X(\zeta) +i\pi \cdot \Pi(\zeta)~]}$.

We have to compute now the next term in the OPE of
these operators as in (\ref{operatorproductofUs}) in order to see
if there are nonzero transitions to the states not
belonging to this family. We have
\beqa
U^{T_1}_{k_1 , \pi_1}(\zeta_1)~ U^{T_1}_{k_2 , \pi_2}(\zeta_2)~\simeq  ~~~~~~~~~~~~~~~~~~~~~~\nonumber\\
(\alpha^2 - \beta^2):
\left(k_1\cdot \partial_{\zeta} X ~ k_1\cdot \partial_{\bar{\zeta}} X -
\pi_1 \cdot \partial_{\zeta} \Pi ~\pi_1 \cdot \partial_{\bar{\zeta}}\Pi -
k_2\cdot \partial_{\zeta} X ~ k_2\cdot \partial_{\bar{\zeta}} X +
\pi_2 \cdot \partial_{\zeta} \Pi ~ \pi_2 \cdot \partial_{\bar{\zeta}}\Pi\right)\nonumber\\
+\alpha\beta ~~\left(k_1\cdot \partial_{\zeta} X ~ k_2\cdot \partial_{\bar{\zeta}} X +
k_2\cdot \partial_{\zeta} X ~k_1\cdot \partial_{\bar{\zeta}} X   +
\pi_1\cdot \partial_{\zeta} \Pi ~ \pi_2\cdot \partial_{\bar{\zeta}} \Pi +
\pi_2\cdot \partial_{\zeta}\Pi ~ \pi_1\cdot \partial_{\bar{\zeta}} \Pi  \right) \nonumber\\
\vert \zeta_1 - \zeta_2 \vert^{2(k_1 \pi_2 + k_2 \pi_1) -2} ~~
W_{k_1 +k_2 , \pi_1 +\pi_2}(\zeta_2): +.....,~~~~
\eeqa
and at the pole $\alpha + \beta =0 $ we get
\beqa
U^{T_1}_{k_1 , \pi_1}(\zeta_1)~ U^{T_1}_{k_2 , \pi_2}(\zeta_2)~\simeq  ~~~~~~~~~~~~~~~~~~~~~~\nonumber\\
\alpha\beta ~~:\left(k_1\cdot \partial_{\zeta} X ~ k_2\cdot \partial_{\bar{\zeta}} X +
k_2\cdot \partial_{\zeta} X ~ k_1\cdot \partial_{\bar{\zeta}} X  +
\pi_1\cdot \partial_{\zeta} \Pi ~ \pi_2\cdot \partial_{\bar{\zeta}} \Pi +
\pi_2\cdot \partial_{\zeta}\Pi ~ \pi_1\cdot \partial_{\bar{\zeta}} \Pi ~ \right) \nonumber\\
\vert \zeta_1 - \zeta_2 \vert^{2(k_1 \pi_2 + k_2 \pi_1) -2} ~~
W_{k_1 +k_2 , \pi_1 +\pi_2}(\zeta_2): +......
\eeqa
The r.h.s. can not be
collected into the $U^{T_{1}}_{k_1 +k_2 , \pi_1 +\pi_2}$ operator. Therefore
the fusion of two $U^{T_1}$'s
of the above form does not scatter to the same state.

The same happens with the second operator $U^{T_2}$, indeed
\beqa
U^{T_2}_{k_1 , \pi_1}(\zeta_1)~ U^{T_2}_{k_2 , \pi_2}(\zeta_2)~\simeq  ~~~~~~~~~~~~~~~~~~~~~~\nonumber\\
\alpha\beta ~~:(k_1\cdot \partial_{\zeta} X ~ k_2\cdot \partial_{\bar{\zeta}} X +
k_2\cdot \partial_{\zeta} X  ~ k_1\cdot \partial_{\bar{\zeta}} X +
\pi_1\cdot \partial_{\zeta} \Pi ~ \pi_2\cdot \partial_{\bar{\zeta}} \Pi +
\pi_2\cdot \partial_{\zeta}\Pi ~ \pi_1\cdot \partial_{\bar{\zeta}} \Pi +\nonumber\\
k_1\cdot \partial_{\zeta} X ~ \pi_2\cdot \partial_{\bar{\zeta}} \Pi +
\pi_2\cdot \partial_{\zeta} \Pi ~ k_1\cdot \partial_{\bar{\zeta}} X  +
\pi_1\cdot \partial_{\zeta} \Pi ~ k_2\cdot \partial_{\bar{\zeta}} X +
k_2\cdot \partial_{\zeta} X  ~ \pi_1\cdot \partial_{\bar{\zeta}} \Pi )\nonumber\\
\vert \zeta_1 - \zeta_2 \vert^{2(k_1 \pi_2 + k_2 \pi_1) -2} ~~
W_{k_1 +k_2 , \pi_1 +\pi_2}(\zeta_2): +.....
\eeqa

\section{{\it Conclusion}}
The vertex operators of the tensionless string model have been
constructed in \cite{alvarez} as polynomials
of the field derivatives $\partial^{m_1}_{\zeta} X^{\mu_1},...,
\partial^{\tilde{m}_j}_{\bar{\zeta}} \Pi^{\tilde{\mu}_j}
e^{~i k \cdot X(\zeta) +i\pi \cdot \Pi(\zeta)~}$ and have anomalous dimension
$\Delta = 2(k\cdot \pi)+ m_1 + \tilde{m}_1 +......+ m_j+ \tilde{m}_j$ .
Of special importance is the vertex operator for fixed helicity states
$$
U_{k,\pi}  = \omega_{\mu\nu } U^{\mu\nu}_{XX}+2\varphi_{\mu\nu } U^{\mu\nu}_{X\Pi}
+\chi_{\mu\nu } U^{\mu\nu}_{\Pi\Pi},
$$
which is quadratic in field derivatives. Its anomalous dimension is
$
\Delta = 2(k \cdot \pi) + 2 =2,
$
that is $(k \cdot \pi) =0$. The last invariant coincides with the length of the
second Casimir operator of the Poincar\'e group $W= (k \cdot \pi)^2 =0$.
Therefore it creates fixed helicity states. All other operators have
$W \neq 0$ and create continuous spin representations - CSR, which are
longitudinal modes in accordance with the conjecture made in \cite{Savvidy:2003fx}.

In order to justify this picture one should prove that all transition
amplitudes to the states of continuous spin representations are equal
to zero and that the only nonzero amplitudes are between fixed helicity states
created by the $U_{k,\pi}$ operator. We don't know the full solution
of this problem, but have been able to derive the conditions
(\ref{polarizationequation1}), (\ref{polarizationequation2}) together
with  (\ref{zeroresiduecondition}) and (\ref{residuepolynomial})
on the operator $U_{k,\pi}$ which guarantee the zero transition
amplitude to the state created by the lowest order CSR operator  $V_{k,\pi}$
in tree approximation.

If the calculation of the conformal dimensions (\ref{conformaldimensions}) for the vertex
operators (\ref{highordervertex}) was for some reason incorrect,
then the conclusion about $V$ and $U$ operators can in principle be changed.
For example, if there is
some undefined source of contribution to the
conformal dimension in (\ref{conformaldimensions}),
then the conclusion indeed can change. Suppose that there is an additional term
to the conformal dimension like in the following formula
$
\Delta = 2(k \cdot \pi)  + m_1 + \tilde{m}_1 +...
...+ m_j+ \tilde{m}_j + 2 ~,
$
then $V$ becomes physical operator and $U$ unphysical because of additional 2.
The Abelian constraint $\Theta$ can be such a source of additional contribution,
through the corresponding Faddeev-Papov ghosts.
But the Faddeev-Popov determinant
is trivial here because there are no derivatives in the
Abelian constraint $\Theta =\Pi^2 -1 =0$.
\vspace{2pt}

Let me also mention that it is generally expected that
the tensionless limit $\alpha^{'} \rightarrow \infty$
of the standard string theory with the Nambu-Goto  area action should have
massless spectrum, because all masses at every level tend to zero as
$M^2_N  = (N-1)/\alpha^{'} \rightarrow 0$\cite{gross}. Of course this
simple conclusion  ignores the importance
of the high genus $g$ diagrams, the contribution of which $A_g$ is exponentially
large compared to the tree level diagram \cite{gross}. The ratio of the corresponding
scattering amplitudes behaves as $A_{g+1} /A_g \simeq exp\{\alpha^{'} s/g^2\}$ and
makes  any perturbative statement unreliable and  requires therefore
nonperturbative treatment of the problem\footnote{The different aspects
and models of tensionless theories can be found in
\cite{deVega:1987hu,Amati:1988tn,Gasperini:1991rv,
Lindstrom:1990qb,DeVega:1992tm,Lizzi:1994rn,Bakas:2004jq}.}.

The tensionless model with perimeter action suggested in  \cite{Savvidy:dv,Savvidy:2003fx,geo}
does not appear as a $\alpha^{'} \rightarrow \infty$ limit of the standard string
theory, as one could probably think,
but has a tensionless character by its geometrical nature \cite{geo}.
Therefore it remains mainly unclear at the moment how these two models are connected.
However the perimeter model shares many properties with
the area strings in the sense that it has
world-sheet conformal invariance, contains the corresponding Virasoro algebra,
which is extended by additional Abelian generators. This makes mathematics
used in the perimeter model very close to the standard string theory and allows
to compute its spectrum \cite{Savvidy:dv,Savvidy:2003fx} and to construct
the appropriate vertex operators \cite{alvarez}.

Comparing literally the spectrum of these two models
one can see that instead of exponential growing of states in the standard
string theory, in the perimeter case we have only linear growing of physical states.
In this respect the number of states in the perimeter model is much less compared with
the standard string theory  and is larger compared with the standard field
theory models of the Yang-Mills type.
From this point of view it is therefore much closes to the quantum field theory rather
than to the standard string theory. At the same time its formulation and
the symmetry structure is more string-theoretical.
Perhaps there should be strong nonperturbative rearrangement of the spectrum
in the limit $\alpha^{'} \rightarrow \infty$ before the spectrum of the area and
the perimeter strings can become close to each other.

I would like to thank  Luis  Alvarez-Gaume, Ignatios  Antoniadis, Lars  Brink
and Kumar Narain for stimulating discussions and
CERN Theory Division for hospitality.

\section{\it {Appendix}}
The analysis of the physical Fock space is based
on the equations which follow from Abelian constraint
$\Theta =\Pi^2 -1 =0$ \cite{Savvidy:2003fx}.
The
$\Pi^{\mu}  = {1\over 2}(\Pi^{\mu}_{L}(\zeta^{+})
+ \Pi^{\mu}_{R}(\zeta^{-}))
$ field is
\beqa\label{decom2}
\Pi^{\mu}_{L}=   e^{\mu} +  k^{\mu}\zeta^{+} +
i\sum_{n \neq 0}  {1\over n }~ \alpha^{\mu}_{n} e^{-in\zeta^{+}},\nonumber\\
\Pi^{\mu}_{R}=   e^{\mu} + k^{\mu}\zeta^{-} +
i\sum_{n \neq 0}  {1\over n }~ \tilde{\alpha}^{\mu}_{n} e^{-in\zeta^{-}}.
\eeqa
and for the lowest excitations it has the form
\be
\Pi^{\mu}(\sigma,\tau)  = e^{\mu} +  k^{\mu} \tau  + {i \over 2} \alpha^{\mu}_{1} e^{-i\zeta^{+}}
-{i\over 2} \alpha^{\mu}_{-1} e^{ i \zeta^{+}} +...
\ee
Thus the constraint equation $\Pi^2 -1 =0$ will take the form
\beqa
\Pi^2 -1  ~~~~~~~~~=~~~~~~ e^2 -1 +  {1 \over 2} \alpha_{1} \alpha_{-1} +\nonumber\\
k^2 ~\tau^2 + 2 (e\cdot k) ~\tau  ~     + \nonumber\\
i (k \cdot  \alpha_{1}) ~\tau ~e^{-i\zeta^{+}} ~~~~-  ~~
i (k \cdot  \alpha_{-1})~ \tau~e^{i\zeta^{+}} ~~+\nonumber\\
i (e \cdot  \alpha_{1})~ e^{-i\zeta^{+}} ~-   ~~~~
i (e \cdot  \alpha_{-1})~ e^{i\zeta^{+}} -\nonumber\\
-{1 \over 4} \alpha_{1} \alpha_{1}e^{-2i\zeta^{+}} -
{1 \over 4} \alpha_{-1} \alpha_{-1}e^{2i\zeta^{+}}
..... =0.
\eeqa
In components it is equivalent to the system of constraint equations
(formulas (32) in \cite{Savvidy:2003fx})
\be
k^2 =0, ~~~~ k\cdot e=0 , ~~~~~ e^2 =1 - {1 \over 2} \alpha_{1} \alpha_{-1},
\ee
\be\label{transfer}
(k \cdot  \alpha_{\pm 1})=0,~~~~(e \cdot  \alpha_{\pm 1})=0.
\ee
\be\label{lightlike}
\alpha_{1} \alpha_{1}=0,~~~~
\alpha_{-1} \alpha_{-1}=0.
\ee
The question is what they mean for the oscillators $\alpha^{\mu}_{\pm 1}$ .

To find an answer let us consider the world-sheet time
$\tau$ evolution of the field $\Pi(\sigma,\tau)$.
We have to fix the initial conditions of the field $\Pi(\sigma,\tau)$.
For that we have to
define the polarization vector $e$, the momentum k and
the coefficients $\alpha_{\pm n}$ in (\ref{decom2}).
Because of the existence of the constraints, the variables e, k,  $\alpha_{\pm n}$ are
not dynamically independent.
The equations (\ref{transfer}) tell us that they are transverse and  longitudinal
at the same time. The equations (\ref{lightlike}) tell us that they are light-like.

The  polarization vector $e$ can be equal to $e_1$ or $e_2$ where
$e_1$ and $e_2$ are two fixed perpendicular $e_1 \cdot e_2 =0$ purely
spatial vectors $-e^{2}_0 + \vec{e}^{2} =1$
(i.e., whose $e_0$ component is 0 and for simplicity we are considering
four-dimensional space-time) which are orthogonal to the fixed light-like momentum vector k
\be
k \cdot e_1 =  k \cdot e_2 =0.
\ee

If we take the initial condition as $(k, e_1)$ then the constraint
equations tell us that the
physical oscillations of $\alpha_{1}$ are those which are perpendicular to k and to $e_1$
$$
(k \cdot  \alpha_{\pm 1})=0,~~~~(e_1 \cdot  \alpha_{\pm 1})=0,
$$
thus
\be
\alpha^{\mu}_{\pm 1} = \alpha_{\pm 1} ~k^{\mu} + B_{\pm}~ e^{\mu}_2 ~,
\ee
and substituting it into the last equation (14) we shall get
$\alpha_{1} \alpha_{1}= B^{2}_+ =  0$, that is
$B_{+} =0$. We have also $\alpha_{-1} \alpha_{-1}= B^{2}_- =  0$, thus $B_{-} =0$.

If we take the initial conditions in the second possible form $(k,e_2)$,
then the constraint equations will tell us that physical oscillations of the
oscillators $\alpha_{\pm 1}$ are in the $e_1$ direction and the oscillations in the $e_2$
direction are forbidden
$$
(k \cdot  \alpha_{\pm 1})=0,~~~~(e_2 \cdot  \alpha_{\pm 1})=0,
$$
thus
\be
\alpha^{\mu}_{\pm 1} = \alpha_{\pm 1} ~k^{\mu} + B_{\pm}~ e^{\mu}_1
\ee
and from the last equation (14) we shall get $B_{\pm} =0$.
Therefore in all cases the oscillators $\alpha^{\mu}_{\pm 1}$  are purely longitudinal
\be
\alpha^{\mu}_{\pm 1} = \alpha_{\pm 1} ~k^{\mu},
\ee
and it is natural to introduce the scalar oscillators
$\alpha_{\pm 1}$.  The $\Pi$ field will take the form
\be
\Pi^{\mu}(\sigma,\tau)  = e^{\mu} +  k^{\mu} \left \{ \tau  + {i \over 2} \alpha_{1} e^{-i\zeta^{+}}
-{i\over 2} \alpha_{-1} e^{ i \zeta^{+}} \right \}+...
\ee
In general case we shall have
\be
\Pi^{\mu}(\sigma,\tau)  = e^{\mu} +  k^{\mu} \left \{ \tau  +
i\sum_{n \neq 0}  {1\over 2n }~ ( \alpha_{n} e^{-in\zeta^{+}} +
~ \tilde{\alpha}^{\mu}_{n} e^{-in\zeta^{-}}) \right \}.
\ee
The last formula has clear physical interpretation: {\it it defines the
polarization vector $e^{\mu}$ up to a large gauge transformation of the from}
\be
e^{\mu} \rightarrow e^{\mu} + k^{\mu} f(\tau,\sigma, a_n)~.
\ee
This was the classical consideration. In the course of covariant quantization
our constraints will be translated into the constraints imposed on the
wave function \cite{Savvidy:2003fx}. Because the
oscillators $\alpha^{\mu}_{\pm n} = k^{\mu} ~\alpha_{\pm n} $
are pure longitudinal, they create only zero norm states.

\vfill

\begin{thebibliography}{99}

\bibitem{Savvidy:dv}
G.~K.~Savvidy,
Phys.Lett.\ B {\bf 552} (2003) 72.


\bibitem{Savvidy:2003fx}
G.~K.~Savvidy,~
``Tensionless strings: Physical Fock space and higher spin fields,''
Int.\ J.\ Mod.\ Phys.\ A {\bf 19},  (2004) 3171-3194;~hep-th/0310085.


\bibitem{geo}G. Savvidy and K. Savvidy. Mod.Phys.Lett. A8 (1993) 2963;
Int.J.Mod.Phys. A8 (1993) 3993; R.V.Ambartzumian and et al, Phys.Lett. B275 (1992) 99;
G. Savvidy. JHEP 0009 (2000) 044;
R.~Manvelian and G.~Savvidy,Phys.Lett.B533 (2002) 138

\bibitem{Antoniadis:2004uh}
I.~Antoniadis and G.~Savvidy,
``Physical Fock space of tensionless strings,''
arXiv:hep-th/0402077.

\bibitem{gross} D.Gross, Phys.Rev.Lett. 60 (1988) 1229

\bibitem{alvarez} L.~ Alvarez-Gaume, I.~ Antoniadis, L.~ Brink, K.~ Narain and G.~ Savvidy
``Tensionless Strings,
Vertex Operators and Scattering Amplitudes '',
Preprint CERN-PH-TH/2004-095 and NRCPS-HE-2004-13, to be published

\bibitem{wigner}E.~Wigner, in {\it Theoretical Physics ed. A.Salam}
(International Atomic Energy, Vienna, 1963) p 59;~
E.~Wigner, Ann. Math. {\bf 40} (1939) 149.

\bibitem{brink}L.~Brink, A.~M.~Khan, P.~Ramond and X.~Xiong,
J. Math. Phys. {\bf 43} (2002) 6279




\bibitem{deVega:1987hu}
H.~J.~de Vega and N.~Sanchez,
Phys.\ Lett.\ B {\bf 197} (1987) 320.


\bibitem{Amati:1988tn}
D.~Amati, M.~Ciafaloni and G.~Veneziano, Phys.\ Lett.\ B {\bf 216} (1989) 41.


\bibitem{Gasperini:1991rv}
M.~Gasperini, N.~Sanchez and G.~Veneziano,
Nucl.\ Phys.\ B {\bf 364} (1991) 365.


\bibitem{Lindstrom:1990qb}
U.~Lindstrom, B.~Sundborg and G.~Theodoridis,
Phys.\ Lett.\ B {\bf 253} (1991) 319.


\bibitem{DeVega:1992tm}
H.~J.~De Vega and A.~Nicolaidis,
Phys.\ Lett.\ B {\bf 295} (1992) 214.


\bibitem{Lizzi:1994rn}
F.~Lizzi,
Mod.\ Phys.\ Lett.\ A {\bf 9}, 1495 (1994)

\bibitem{Bakas:2004jq}
I.~Bakas and C.~Sourdis,
``On the tensionless limit of gauged WZW models,''
arXiv:hep-th/0403165.

\bibitem{pauli}M.~Fierz and W.~Pauli,
Proc.Roy.Soc. A173 (1939) 211\\
W.Rarita and J.Schwinger, Phys.Rev. {\bf 60} (1941) 61\\
J.Schwinger, {\it Particles, Sourses, and Fields}
(Addison-Wesley, Reading, MA, 1970)


\bibitem{singh}L.~P.~S.~Singh and C.~R.~Hagen,
Phys.Rev. {\bf D9} (1974) 898, 910

\bibitem{fronsdal}C.Fronsdal,
Massless fields with integer spin,
Phys.Rev. {\bf D18} (1978) 3624


\bibitem{Bengtsson:1983pg}
A.~K.~Bengtsson, I.~Bengtsson and L.~Brink,
Nucl.\ Phys.\ B {\bf 227} (1983) 41;Nucl.\ Phys.\ B {\bf 227} (1983) 31.




\end{thebibliography}
\end{document}